\documentclass[twocolumn, 10pt, pra, aps, showpacs, reprint, amsmath, amssymb, superscriptaddress, nofootinbib]{revtex4-2}

\usepackage[]{graphicx}

\begin{document}

\title{M\"obius Strip Microlasers: a Testbed for Non-Euclidean Photonics}

\author{Yalei Song}
\affiliation{Laboratoire Lumi\`ere, Mati\`ere et Interfaces (LuMIn)
CNRS, ENS Paris-Saclay, Universit\'e Paris-Saclay, CentraleSup\'elec, 91190 Gif-sur-Yvette, France.}
\affiliation{Lanzhou Center for Theoretical Physics and the Gansu Provincial Key Laboratory of Theoretical Physics, Lanzhou University, Lanzhou, 730000 Gansu, China.}
\author{Yann Monceaux}
\affiliation{Laboratoire Lumi\`ere, Mati\`ere et Interfaces (LuMIn)
CNRS, ENS Paris-Saclay, Universit\'e Paris-Saclay, CentraleSup\'elec, 91190 Gif-sur-Yvette, France.}
\author{Stefan Bittner}
\affiliation{Chair in Photonics, LMOPS EA-4423 Laboratory, CentraleSup\'elec and Universit\'e de Lorraine, 2 rue Edouard Belin, 57070 Metz, France}
\author{Kimhong Chao}
\affiliation{Laboratoire Lumi\`ere, Mati\`ere et Interfaces (LuMIn)
CNRS, ENS Paris-Saclay, Universit\'e Paris-Saclay, CentraleSup\'elec, 91190 Gif-sur-Yvette, France.}
\author{H\'ector M. Reynoso de la Cruz}
\affiliation{Laboratoire Lumi\`ere, Mati\`ere et Interfaces (LuMIn)
CNRS, ENS Paris-Saclay, Universit\'e Paris-Saclay, CentraleSup\'elec, 91190 Gif-sur-Yvette, France.}
\affiliation{Department of Physical Engineering, Academic Body of Statistical Mechanics, Science and Engineering Division of the University of Guanajuato, Le\'on, Gto. 36000, M\'exico}
\author{Cl\'ement Lafargue}
\affiliation{Laboratoire Lumi\`ere, Mati\`ere et Interfaces (LuMIn)
CNRS, ENS Paris-Saclay, Universit\'e Paris-Saclay, CentraleSup\'elec, 91190 Gif-sur-Yvette, France.}
\author{Dominique Decanini}
\affiliation{Centre de Nanosciences et de Nanotechnologies, CNRS, Universit\'e Paris-Saclay, 10 Boulevard Thomas Gobert, 91120 Palaiseau, France.}
\author{Barbara Dietz}
\affiliation{Lanzhou Center for Theoretical Physics and the Gansu Provincial Key Laboratory of Theoretical Physics, Lanzhou University, Lanzhou, 730000 Gansu, China.}
\author{Joseph Zyss}
\affiliation{Laboratoire Lumi\`ere, Mati\`ere et Interfaces (LuMIn)
CNRS, ENS Paris-Saclay, Universit\'e Paris-Saclay, CentraleSup\'elec, 91190 Gif-sur-Yvette, France.}
\author{Alain Grigis}
\affiliation{Laboratoire d'Analyse, G\'eom\'etrie et Applications, CNRS UMR 7539, Universit\'e Sorbonne Paris Cit\'e, Universit\'e Paris 13, Institut Galil\'ee, 99 avenue Jean-Baptiste Cl\'ement, 93430 Villetaneuse, France}
\author{Xavier Checoury}
\affiliation{Centre de Nanosciences et de Nanotechnologies, CNRS, Universit\'e Paris-Saclay, 10 Boulevard Thomas Gobert, 91120 Palaiseau, France.}
\author{Melanie Lebental}
\affiliation{Laboratoire Lumi\`ere, Mati\`ere et Interfaces (LuMIn)
CNRS, ENS Paris-Saclay, Universit\'e Paris-Saclay, CentraleSup\'elec, 91190 Gif-sur-Yvette, France.}
\email{lebental@ens-paris-saclay.fr}

\date{\today}

\begin{abstract}
We report on experiments with M\"obius strip microlasers which were fabricated with high optical quality by direct laser writing. A M\"obius strip, i.e., a band with a half twist, exhibits the fascinating property that it has a single nonorientable surface and a single boundary. We provide evidence that, in contrast to conventional ring or disk resonators, a M\"obius strip cavity cannot sustain whispering gallery modes (WGM). Comparison between experiments and 3D finite difference time domain (FDTD) simulations reveals that the resonances are localized on periodic geodesics.
\end{abstract}

\maketitle

The search for geodesic curves on manifolds of arbitrary dimension and metric has been a major driving force in mathematical physics from its inception at the beginning of the 20th century \cite{hadamard,poincare,levi-civita}. The classical work by J.~L.\ Synge brought this domain in contact with general relativity \cite{synge}. In the 1970s, ``trace formulas'' were developed that provide a semiclassical approximation of the spectral density in terms of a sum over classical periodic trajectories, establishing a direct link between classical and quantum mechanics \cite{brack,stoeckmann-livre}. The seminal works of Balian \cite{balian} and Gutzwiller \cite{gutzwiller} were implemented in a broad diversity of physical systems, including electron transport \cite{electron-transport}, billiards \cite{berry}, and nuclear physics \cite{patricio}. For open wave systems, the modes are typically localized on certain classical trajectories \cite{harayama}. Similarly, the motion of particles constrained to a compact Riemannian surface \cite{prb-goldstone,dacosta} of negative curvature is connected to classical geodesics via the Selberg trace formula \cite{gutzwiller,sieber-hyperbolique}. 

Variational principles are of fundamental importance in physics and also at the origin of the trace formula. The stationary phase approximation relates the quantum (or wave) propagator to classical trajectories, which follow the principle of least action \cite{brack}. On curved surfaces, these trajectories are not straight lines but \emph{geodesics}, which are the shortest paths between two points \cite{Kuehnel}. Surfaces of arbitrary local curvature provide a fascinating playground for non-Euclidean optics. Only recently have advanced lithography technologies enabled the fabrication of three-dimensional (3D) surfaces with high optical quality and thicknesses down to $1~\mu$m \cite{2PP}, confining light to propagation within a quasi-two-dimensional curved layer. Many open questions can be tackled regarding the structure of the electromagnetic field in non-Euclidean resonators \cite{epl-hentschel,peschel-PRL,arie-polariton,sebbah}, which are also of relevance in other wave-related research fields like acoustics \cite{acoustic-3D}, hydrodynamics, gravity, or quantum physics. 

\begin{figure}[tb]
\begin{center}
\includegraphics[width = 0.8\linewidth]{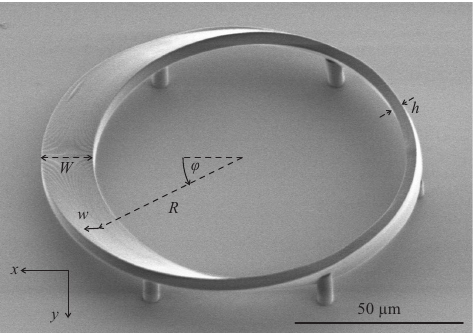}
\end{center}
\caption{SEM image of a M\"obius strip microlaser with radius $R=50\,\mu$m, width $W=15\,\mu$m, and thickness $h=3\,\mu$m. }
\label{fig:SEM}
\end{figure}

Some aspects of quantum-mechanical systems can be investigated with electromagnetic systems thanks to the equivalence of the Schr\"odinger and Helmholtz equations. Prominent examples include billiard systems and their experimental implementations with microwave resonators \cite{stoeckmann-livre}, dielectric resonators \cite{bittner1, bittner2, cao-review}, and microlasers \cite{PRE-trace1, PRE-trace2}, with ramification extending from applied to mathematical physics \cite{dietz, bellec2013, Rechtsman2013}. Organic microlasers provide an ideal testbed for studying ray-wave correspondence: with a typical size in the range of $50$--$200~\mu$m, much larger than the wavelength, they operate in the semiclassical regime. 

\begin{figure}[tb]
\includegraphics[width=.99\linewidth]{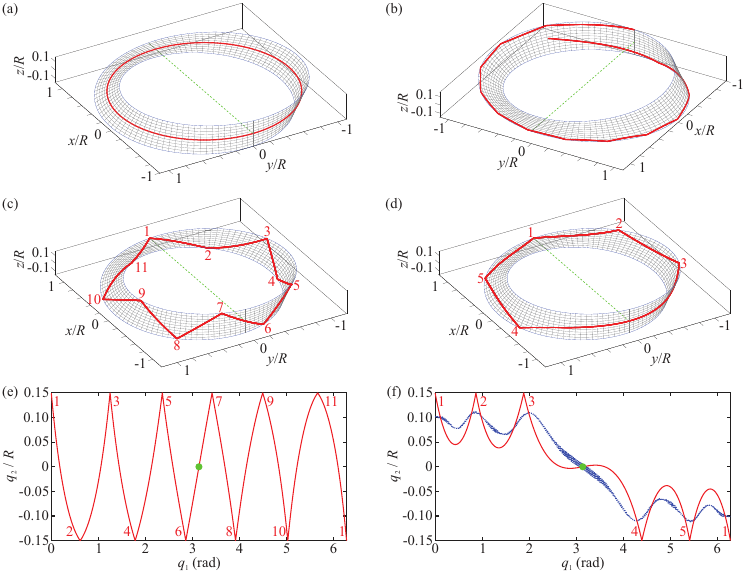}
\caption{Examples of trajectories on a M\"obius strip with $W/R$ = 0.3. The rotation symmetry axis $y$ = $z$ = 0 is indicated in green (dashed lines or dots). (a) Path along the center line of the M\"obius strip. It is \textit{not} a geodesic. (b) Trajectory along the boundary of the M\"obius strip. It has to cross over to the ``opposite'' side at concave boundary parts. (c),(e) A periodic geodesic with 11 vertices and length $L = 7.031 R$. (d),(f) Periodic geodesic 5a with five vertices and length $L = 6.732 R$. The blue dotted line is the mean position calculated from Eq.~(\ref{eq:mean-position}) for the wave function in Fig.~\ref{fig:num-resonance}.}
\label{fig:geodesique}
\end{figure}

In this Letter, we explore the emerging domain of non-Euclidean photonics with M\"obius strip microlasers fabricated by direct laser writing, see Fig.~\ref{fig:SEM}. M\"obius strips have captured the attention of generations of scientists, from pure mathematics to physical, chemical and engineering sciences \cite{nature-cristal,isolant-topo,prb-electron,lpr-plasmon,epl-hentschel,plasmon-chine}. A M\"obius strip is formed by connecting two ends of a strip after twisting one end by 180$^{\circ}$, whereas connecting them without twist results in a normal ring resonator.
In spite of its simplicity, it exhibits peculiar topological properties because it has a single, nonorientable surface and a single boundary. In particular, a M\"obius microcavity cannot display whispering gallery modes (WGMs). Normal ring resonators feature high-$Q$ WGMs that propagate along the outer circular boundary. The trace formula relates WGMs to periodic ray orbits in the form of regular polygons with reflections at the outer boundary  \cite{brack-circle,bittner1}. However, such classical trajectories do not exist in a M\"obius strip because its boundary exhibits both concave and convex parts, see Fig.~\ref{fig:geodesique}(b). We provide evidence that, instead, its low-loss resonances are localized on periodic geodesics. 

\section{M\"obius fabrication and parametrization}
The M\"obius strip was designed by rotating a rectangle by 180$^{\circ}$ while its center travels along a circle of radius $R$. The geometric parameters are explained in Fig.~\ref{fig:SEM}. The width is fixed $W=15~\mu$m. The radius $R$ varies from $40$ to $60~\mu$m, and the thickness $h$ varies from $1$ to $5~\mu$m.

The microlasers were fabricated by direct laser writing lithography using a Photonic Professional GT system with negative resist IP-G780 from the Nanoscribe company \cite{2PP}. The resist was doped by $0.5$ wt\% pyrromethene 597 laser dye (by the company Exciton), which is homogeneously distributed in the bulk host resist \cite{pyramid}. Each cavity is supported by six circular pylons to avoid coupling with the glass substrate. A scanning electron microscope (SEM) image of a M\"obius strip microlaser is shown in Fig.~\ref{fig:SEM}. 

A standard parametrization of a M\"obius strip surface $\vec r=\vec F(\varphi,w)$ is
\begin{eqnarray}
x&=&F_x(\varphi,w)=(R+w\cos\frac{\varphi}{2})\,\cos\varphi\label{eq:moebiusx} \\
y&=&F_y(\varphi,w)=(R+w\cos\frac{\varphi}{2})\,\sin\varphi\label{eq:moebiusy} \\
z&=&F_z(\varphi,w)=w\sin\frac{\varphi}{2}\label{eq:moebiusz}
\end{eqnarray}
with $w\in[-W/2;W/2]$ and $\varphi\in[0;2\pi]$, which yields a M\"obius strip like in Fig.~\ref{fig:SEM}. A M\"obius strip can be twisted the other way, which corresponds to adding a minus sign on line~(\ref{eq:moebiusy}). We checked that the experiments and their interpretation are consistent for both chiralities.

Note that this parametrization differs from a M\"obius strip constructed from a twisted paper sheet \cite{Gravesen}. Since the latter can be unfolded to a flat plane, we call it a \emph{flat} M\"obius strip, see Appendix~\ref{sec:FlatMoebius}. This has important consequences for geodesics and mode localization, which will be discussed below.

The M\"obius strip microlasers were pumped individually and with uniform intensity by a beam perpendicular to the substrate from a frequency doubled Nd:YAG laser (532 nm, 500 ps, 10 Hz). Their emission was analyzed by a spectrometer coupled to a CCD camera with an overall resolution of 0.03 nm. The experiments were carried out at room pressure and temperature.

\begin{figure}[tb]
\begin{center}
\includegraphics[width = 0.9\linewidth]{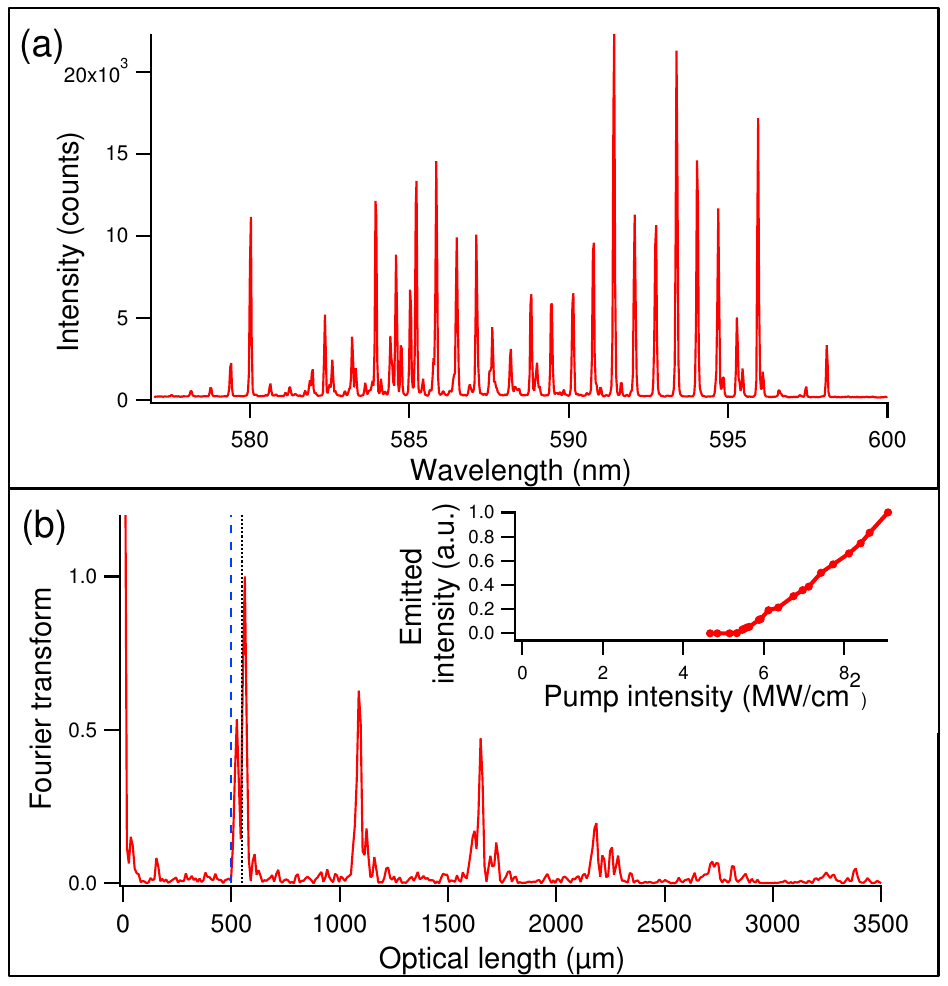}
\end{center}
\caption{(a) Experimental laser spectrum of a M\"obius strip with $R=50$ $\mu$m and $h=1$ $\mu$m ($1$~s exposure). (b) Normalized Fourier transform (from wave number $k$ to optical length) of the spectrum in (a). The vertical lines are at the optical lengths of half the perimeter (dashed blue line) and of the geodesic 5a (black dotted line) for $n_g=1.58$. Inset: experimental threshold curve for a M\"obius microcavity with $R=50$ $\mu$m and $h=3~\mu$m.}
\label{fig:manips-spectre}
\end{figure}

The inset of Fig.~\ref{fig:manips-spectre} shows a laser threshold at $5$~MW/cm$^2$. This value decreases with increasing size of the M\"obius strip, because of a larger gain volume collecting more pump light, and is similar to the laser threshold of 3D cavities made of the same laser dye with an equivalent gain volume \cite{pyramid}.

The microlaser light is mostly emitted parallel to the substrate plane, and the emission mainly originates from the lateral boundaries of the strip at a near-grazing angle, see Fig.~\ref{fig:photo-pompage} in Appendix~\ref{sec:addExp}. This is observed from all directions, which is consistent with modes confined inside the M\"obius strip and propagating within this curved and twisted guiding layer.

A laser emission spectrum is shown in Fig.~\ref{fig:manips-spectre}(a). It does not significantly depend on the direction of observation, but depends on the thickness $h$ as shown in Appendix~\ref{ssec:thickness}. It features a single series of equidistant resonances for $h=1~\mu$m, which reveals information on the mode localization.

\section{Helmholtz equation}
In Cartesian coordinates, the electric field fulfills the Helmholtz equation
\begin{equation}\label{eq:helmholtz-cart}
(\Delta + n^2k^2)\vec E=\vec 0
\end{equation}
where $k$ is the wave number in vacuum and $n$ is the refractive index, which is $1$ outside of the strip and $n\simeq 1.5$ inside it \cite{indice}. As the fabrication method does not generate stress in the resist, $n$ is assumed to be homogeneous within the strip.
Although it is experimentally evidenced that the laser emission of a M\"obius microcavity is polarized, for the sake of simplicity, we will only consider a single electric field component $\psi$.

To map the Helmholtz equation (\ref{eq:helmholtz-cart}) onto the M\"obius surface, we introduce a local coordinate system ($q_1$,$q_2$,$q_3$) defined by the local frame ($\vec u_1$,$\vec u_2$,$\vec u_3$) with
\begin{equation}\label{eq:coordonnees-curvilignes}
\vec u_1=\frac{\partial\vec F}{\partial \varphi}\hspace{0.5cm}\vec u_2=\frac{\partial\vec F}{\partial w}\hspace{0.5cm}\vec u_3=\frac{\vec u_1\times\vec u_2}{||\vec u_1\times\vec u_2||} \, .
\end{equation}
By construction, $\vec u_1$ and $\vec u_2$ lie within the strip, whereas $\vec u_3$ is orthogonal to it. For the M\"obius strip considered here $(\varphi,w)=(q_1,q_2)$.

We deal with the finite, but small thickness ($h = 1~\mu$m) of the M\"obius strip by introducing an effective refractive index\footnote{Another approach would be to add a $q_3$ component to the parametrization equations~(\ref{eq:moebiusx})--(\ref{eq:moebiusz}), then to change from the Cartesian coordinate to the local coordinate system, and finally to shrink the thickness to zero or to a small $h$ value. In Ref.~\cite{dacosta}, this method was employed for matter waves. For electromagnetic waves, the effective index approximation has been widely and successfully used for systems composed of flat dielectric layers.}, then we assume that the direction $q_3$ perpendicular to the strip can be separated from the directions $(q_1, q_2)$ within the strip. Solving the equation in the $q_3$ direction yields the effective index for the phase velocity of a wave guided within the M\"obius strip. We extend the effective index $n_\mathrm{eff}$, originally introduced for flat layers, to a curved layer following Ref.~\cite{jap-indice-courbe}. The derivation is described in Appendix~\ref{ssec:derivation}, where it is shown that the correction due to the curvature is negligible for the range of parameters considered in the experiments. 

With these approximations, the wave equation reduces to a two-dimensional Helmholtz equation within the strip,
\begin{equation}
\Delta_s\psi_s+n_\mathrm{eff}^2k^2\psi_s=0 \, ,
\end{equation}
where $\psi_s$ depends only on $(q_1, q_2)$. The Laplace operator $\Delta_s$ in the curvilinear coordinate system is given by
\begin{equation}
\Delta_s \psi=\frac{1}{\sqrt g}\sum_{i,j=1}^2\frac{\partial}{\partial q_i}\left(
\sqrt{g}\,g^{ij}\frac{\partial \psi}{\partial q_j}\right)
\end{equation}
with $g_{ij}=(\vec u_i\,\cdot \,\vec u_j)$ as the metric tensor, $(g^{ij})$ as its inverse, and $g=\det (g_{ij})$. 

Based on this approximation\footnote{The surface parametrization equations~(\ref{eq:moebiusx})--(\ref{eq:moebiusz}) does not depend on $q_3$. It follows that $g_{33}=1$ and $g_{13}=g_{31}=g_{23}=g_{32}=0$, and that the 3D curvilinear Laplace operator can be separated in two parts, namely the propagation within the curved surface, and the propagation orthogonal to the strip. In practice, the actual M\"obius cavity has a nonvanishing thickness $h$, and both directions of propagation are coupled. It is assumed that this coupling is negligible for a small thickness, thus allowing us to introduce the effective refractive index $n_{\textrm{eff}}$.}, we consider the propagation of rays with an effective index $n_{\textrm{eff}}$ in the curved surface defined by $(\vec u_1,\vec u_2)$. Since the radius $R$ and the width $W$ of the strip are large compared to the wavelength, a semiclassical approach is valid. The resonant modes should hence be related to periodic geodesics.

\section{Geodesics}
A classical trajectory $\gamma$ on a surface can be represented by a path $q(s)=[q_1(s),q_2(s)]$ with $s\in[0,1]$; $q(0)$ is the initial point and $q(1)$ is the final point. This trajectory $\gamma$ is a geodesic if its length is stationary with respect to variations of the path $q(s)$ while maintaining $q(0)$ and $q(1)$. The length is given by \cite{Kuehnel}
\begin{equation} \label{eq:Lgamma}
L_{\gamma}= \int_0^1 ds \left[ \sum \limits_{i,j} g_{ij}[q(s)] \,\frac{d q_i}{ds}(s) \,\frac{d q_j}{ds}(s)  \right]^{1/2}
\end{equation}
with $i,j=1,2$. Since the geodesics on the M\"obius strip cannot be derived analytically, we calculated them numerically using this variational principle (see Appendix~\ref{ssec:calcGeodVar}). 

A geodesic is periodic if $q(0)=q(1)$ and $dq/ds(0) = dq/ds(1)$. In contrast to the flat M\"obius billiard (see Appendix~\ref{sec:FlatMoebius}), all periodic geodesics have at least one reflection on the boundary and are isolated\footnote{A periodic geodesic is called isolated if there is no other periodic geodesic in its vicinity.}. Another striking difference is the sequence of reflections on the boundary. In the flat M\"obius billiard, the geodesics have consecutive reflections on opposite sides\footnote{The notion of same or opposite side makes sense from a local perspective only, since a M\"obius strip actually has only one side.}; that is, a reflection with $q_2=+W/2$ is always followed by one with $q_2=-W/2$ and vice versa. Such periodic geodesics exist in the M\"obius strip as well, but only if they have at least 11 reflections, see Fig.~\ref{fig:geodesique}(c). As illustrated in Fig.~\ref{fig:geodesique}(d), it also hosts periodic geodesics with consecutive reflections on the same side, a feature specific to the 3D M\"obius strip. We call it ``geodesic 5a'' to distinguish it from other periodic geodesics with five reflections and similar length. Hereafter, we will compare it to experiments and numerical simulations.

\section{Spectrum analysis}
The equidistant resonances in Fig.~\ref{fig:manips-spectre}(a) indicate that the lasing modes are longitudinal modes located along a single classical trajectory or a few trajectories with similar lengths. To verify this conjecture, we analyze the spectra. 

The distance $\Delta k$ between two adjacent peaks is related to the geometric length $L$ of this classical trajectory via $\Delta k= 2\pi/(n_g L)$ where $n_g$ is the group refractive index \cite{APL-FP3D}. $\Delta k$ can be inferred from the Fourier transform, as illustrated in Fig.~\ref{fig:manips-spectre}(b), which shows peaks at the optical length $n_g L$ and its harmonics.
Hence we can determine the geometric length $L$ if the group refractive index $n_g$ is known. It is deduced from the effective index and contributions from the material dispersion of $n$, and modal dispersion of $n_{\mathrm{eff}}$. The three terms and their uncertainties are described in the Appendix~\ref{ssec:exp} and yield a group index of $n_g=1.58\,\pm\,0.05$. 

\begin{figure}[tb]
\begin{center}
\includegraphics[width = 1\linewidth]{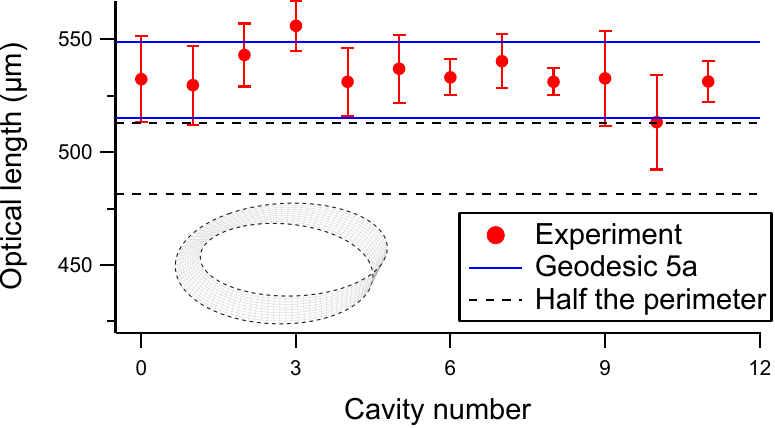}
\end{center}
\caption{Comparison of the experimental optical lengths for 12 M\"obius microlasers with radius $R=50\,\mu$m and thickness $h=1\,\mu$m (red dots) with the optical length of geodesic 5a (continuous blue lines) and of the half perimeter (dashed black lines). The lower and upper lines correspond to the lower ($n_g=1.53$) and upper  ($n_g=1.63$) limit of the group index $n_g$. Inset: the dashed line indicates the perimeter.}
\label{fig:manips}
\end{figure}

Figure~\ref{fig:manips} shows the measured optical lengths for 12 M\"obius microlasers with radius $R=50~\mu$m and thickness $h=1~\mu$m. Their clear disagreement with half the perimeter evidences that the lasing modes do not propagate along the cavity boundary as WGMs would do. In contrast, the agreement is very good for the geodesic 5a. A few other geodesics are also consistent with the experimental data. Some geodesics can be excluded because their length is too short or too long. This includes, for instance, all the geodesics with consecutive reflections on opposite sides, like the one in Fig.~\ref{fig:geodesique}(c).

\section{Wave functions}
\begin{figure}[tb]
\begin{center}
\includegraphics[width = 1\linewidth]{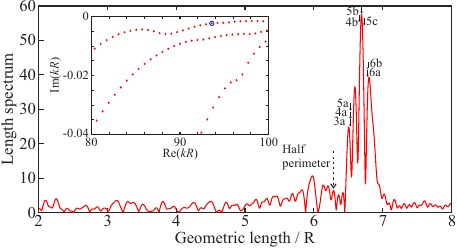}
\end{center}
\caption{Length spectrum derived from FDTD simulations for a M\"obius strip with $R = 10$ $\mu$m, $W = 3~\mu$m, $h = 150$~nm, and $n = 1.515$. The dashed arrow indicates half the length of the perimeter. The solid arrows indicate the lengths of several periodic geodesics with $4$--$6$ reflections. The inset shows part of the simulated spectrum, where the blue circle indicates the mode plotted in Fig.~\ref{fig:num-resonance}.}
\label{fig:num-spectre}
\end{figure}

Numerical simulations were performed with a home-made 3D finite difference time domain (FDTD) code for $R=10~\mu$m, $W=3~\mu$m, $h=150$~nm, and $n = 1.515$. The aspect ratio $W/R$ is the same as in the experiments. For these parameters, there is a single excitation in the vertical $q_3$ direction but there exist several modes in the transverse $w$ direction. The spectrum plotted in the inset of Fig.~\ref{fig:num-spectre} features several branches of equidistant resonances. The $Q$ factors of the low-loss modes (upper branch) are as high as 200,000. The spectrum is shown on a broader range in Fig.~\ref{fig:SpectrumFDTD} in Appendix~\ref{sec:3dFDTD}. 

\begin{figure}[tb]
\begin{center}
\includegraphics[width = 1\linewidth]{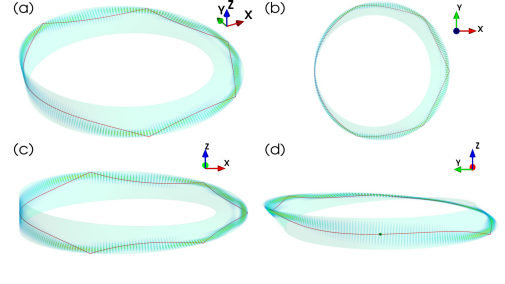}
\end{center}
\caption{Different views of the same mode with $\mathrm{Re}(kR)=93.589$ and $\mathrm{Im}(kR)=-0.0023$ ($Q = 20,345$). $|\vec E|^2$ is shown in false colors. The geodesic 5a is superimposed as red lines. The green dot in (d) indicates the intersection of the $x$ axis with the vertical section of the M\"obius strip.}
\label{fig:num-resonance}
\end{figure}

A typical wave function is plotted in Fig.~\ref{fig:num-resonance}. Because of the two-fold rotational symmetry of the M\"obius strip with respect to the $x$ axis, the wave function must be odd or even with respect to rotation by $180^\circ$ about it, i.e., about the green dot in Fig.~\ref{fig:num-resonance}(d). Consequently, the wave must cross from the upper side of the strip to the lower side, in contrast to a WGM, which would evolve along the boundary. 

This wave function exhibits maximal intensity close to the boundary at positions coinciding with the reflections of the geodesic 5a which is shown as a red line. To quantify the agreement, the mean position of the wave function along the strip width is calculated for each $\varphi$ value,
\begin{equation}
\langle w\rangle(\varphi)=\frac{\iint\limits_{\textrm{Half-plane}} w\,\rho(w,\varphi,q_3)\,dw\,dq_3}
{\iint\limits_{\textrm{Half-plane}}\,\rho(w,\varphi,q_3)\,dw\,dq_3 \, ,}
\label{eq:mean-position}
\end{equation}
where $\rho=\frac{1}{2}\epsilon_0\epsilon_r|\vec E|^2+\frac{1}{2}\mu_0|\vec H|^2$ is the energy density of the electromagnetic field. This mean position is plotted in Fig.~\ref{fig:geodesique}(f). The agreement is very good, in particular for the reflections on the boundary, further corroborating our claim that the modes are localized on periodic geodesics. More examples of wave functions are presented in Figs.~\ref{fig:mode3} and \ref{fig:mode8} in Appendix~\ref{ssec:modes}. 

The length spectrum in Fig.~\ref{fig:num-spectre} is a Fourier transform of the simulated spectrum; see Appendix~\ref{ssec:Lspect} for more information. It is peaked at the lengths of the periodic geodesics that were identified from the experiments and the wave functions, whereas there is no significant peak at the position of the half perimeter. 

\section{Conclusion}
Organic M\"obius strip microlasers are investigated as nontrivial examples of non-Euclidean photonic structures fabricated by direct laser writing. We show that M\"obius strip microlasers do not exhibit WGMs, in contrast to conventional ring cavities. Instead their lasing modes are located on periodic geodesics. These findings are based on experiments, 3D FDTD numerical simulations, and on a dedicated algorithm to systematically identify the periodic geodesics. Our analysis is based on an effective index approximation that reduces the electromagnetic wave equation to a two-dimensional scalar Helmholtz equation. A future objective will be the derivation of a vectorial equation for the modes in curved surfaces and an in-depth investigation of the nontrivial polarization features \cite{bauer-science} of such microlasers. This work opens the way to further explorations of non-Euclidean photonic devices.

\begin{acknowledgments}
This work was supported by the French RENATECH network, the CNano IdF DIM Nano-K, the Labex NanoSaclay (ANR-10-LABX-0035), and the Laboratoire International Associ\'e ImagiNano. Y.\ S.\ and B.\ D.\ are supported by the National Natural Science Foundation of China under Grants No. 11775100 and No. 11775101. Furthermore, B.\ D.\ would like to thank the ENS Paris-Saclay for financial support through MONABIPHOT Erasmus Mundus Master Course program during her stay at the ENS. S.\ B.\ acknowledges support for the Chair in Photonics from Minist\`ere de l'Enseignement Sup\'erieur, de la Recherche et de l'Innovation, R\'egion Grand-Est, D\'epartement Moselle, European Regional Development Fund (ERDF), Metz M\'etropole, GDI Simulation, CentraleSup\'elec, and Fondation CentraleSup\'elec. H.\ M.\ R.\ is supported by Consejo Nacional de Ciencia y Tecnolog\'ia - M\'exico under Grant No. 2019-000016-01NACF-03530 of Movilidad Extranjera 2019, and the Direcci\'on de Apoyo a la Investigaci\'on y el Posgrado (DAIP) Guanajuato under the Convocatoria de apoyo a posgrados-2019 Formato-B. J.-F.\ Audibert and H.\ K.\ Warner are acknowledged for the fabrication of centimeter-scale M\"obius strips with a 3D printer, N.\ Pavloff for pointing out Ref.~\cite{prb-goldstone}, P.~Pansu, F.~Jean, and G.~Bossard for advice on geodesics, and E.~Bogomolny for his derivation of the effective index approximation.
\end{acknowledgments}

\appendix

\section{The flat M\"obius billiard} \label{sec:FlatMoebius}
The M\"obius strip, parametrized by the set of Equations~(\ref{eq:moebiusx})--(\ref{eq:moebiusz}) in the main text, is hereafter referred to as \emph{3D M\"obius strip} to distinguish it from the \emph{flat M\"obius billiard}. Please note that all calculations presented in the appendices are made for a M\"obius strip with chirality corresponding to an additional minus sign in Eq.~(\ref{eq:moebiusy}) for the $y$ coordinate. 

In this section we discuss the flat M\"obius billiard, which has fundamentally different ray dynamics compared to the three-dimensional (3D) M\"obius strip considered in the main text. These differences are discussed at the end of this Appendix, and the structure of the periodic geodesics of the 3D M\"obius strip is compared to those of the flat M\"obius billiard in Appendix~\ref{sec:PG-3Dmoebius}. 

\begin{figure*}[tb]
\begin{center}
\includegraphics[width = 12 cm]{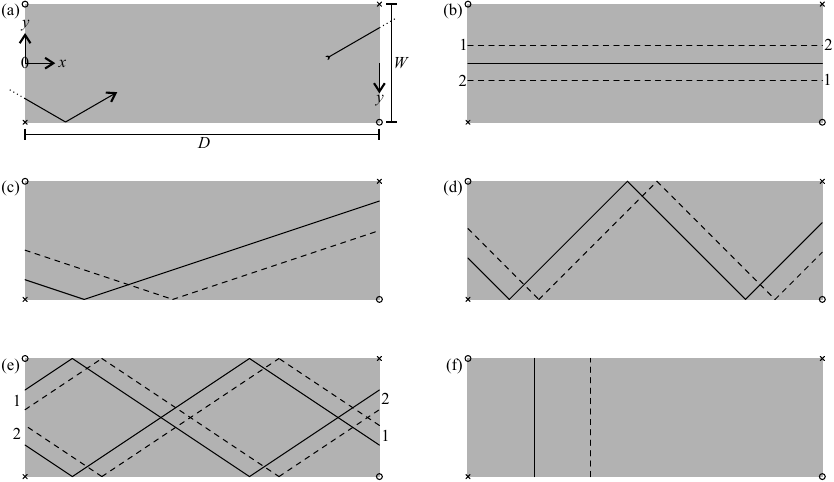}
\end{center}
\caption{\label{fig:POplaneMoebius}
Periodic orbits in the flat M\"obius billiard. (a) Geometry of the flat M\"obius billiard of length $D$ and width $W$. The boundaries at $x = 0$ and $x = D$ are identified with each other after an inversion of the $y$ direction, hence the points marked by o and x are identified with each other, respectively. (b) The isolated periodic orbit $(n_t, n_l) = (0, 1)$ (solid line) and one PO of the $(0, 2)$ family (dashed line). (c) Two examples of the $(1, 1)$ PO family. (d) Two examples of the $(3, 1)$ family. (e) Two examples of the $(4, 2)$ family. (f) Two examples of the $(2, 0)$ PO family, also called bouncing-ball orbits. The numbers $1$ and $2$ in panels (b) and (e) indicate the connections of the different PO segments.}
\end{figure*}

The flat M\"obius billiard is defined as a rectangle with length $D = 2 \pi R$ and width $W$ where the transverse boundaries, defined by $x\in [0, D]$ at $y = \pm W/2$, are hard walls with specular reflections. The diagonally opposite corners are identified with each other, that is, periodic boundary conditions in longitudinal ($x$) direction with an inversion of the transverse ($y$) coordinate are imposed to obtain the topology of a M\"obius strip, that is
\begin{equation} \begin{array}{lcl} x = D & \rightarrow & x = 0 \\ y|_{x=0} & = & -y|_{x=D} \\ \dot{y}|_{x=0} & = & -\dot{y}|_{x=D} \end{array} \end{equation}
as illustrated in Fig.~\ref{fig:POplaneMoebius}(a).

\subsection{Periodic orbits of the flat M\"obius billiard}
The Periodic Orbits (POs) of the flat M\"obius billiard are classified by the index pair $(n_t, n_l)$ where $n_t$ is the number of reflections at the transverse boundaries and $n_l$ is the number of roundtrips in longitudinal ($x$) direction. Several examples are shown in Fig.~\ref{fig:POplaneMoebius}. Since the transverse velocity component $\dot{y}$ must return to its initial value after one roundtrip along a PO, and each reflection as well as each roundtrip in longitudinal direction change the sign of $\dot{y}$, the sum of $n_t$ and $n_l$ must be even. The only exception is the $(0, 1)$ periodic orbit (and its odd repetitions) because $\dot{y}=0$ in this case [see Fig.~\ref{fig:POplaneMoebius}(b)]. It is an isolated PO whereas all other POs are part of a family that covers the whole billiard. 

The lengths of the POs are given by
\begin{equation} L(n_t, n_l) = [ (n_t W)^2 + (n_l D)^2 ]^{1/2} \end{equation}
and their momentum vectors $\vec{k} = (k_x, k_y)$ fulfill $|k_x| = k \cos(\alpha)$ and $|k_y| = k \sin(\alpha)$ with
\begin{equation} \label{eq:POdirection} \alpha(n_t, n_l) = \arctan[n_t W / (n_l D)] \, . \end{equation}
Any trajectory with a momentum vector fulfilling Eq.~(\ref{eq:POdirection}) is a PO, independent of its initial position.

\subsection{Developable and nondevelopable surfaces}
The flat M\"obius billiard is a representation of a developable M\"obius strip. Developable surfaces are smooth surfaces with vanishing Gaussian curvature \cite{Kuehnel}, where the Gaussian curvature is the product of the two principal curvatures at a given point. Developable surfaces have a flat metric like the Euclidean plane, that is, there exists a parametrization of the surface such that the metric tensor is $g_{ij} = \delta_{ij}$. Developable surfaces can be obtained by rolling a piece of paper without stretching, compression, folding or similar distortions. A cylinder is an example of a developable surface, but also a M\"obius created from a paper strip. Since the geodesics of a surface only depend on its metric, the flat M\"obius billiard considered above is equivalent to any developable M\"obius strip (like the one in Ref.~\cite{Gravesen}) as far as their ray dynamics is concerned. 

Nondevelopable surfaces, on the other hand, have a nonvanishing Gaussian curvature and a metric that is not flat. This is the case for the 3D M\"obius strip considered here. The differences in the structure of the periodic geodesics and the ray dynamics in general between the flat M\"obius billiard and the 3D M\"obius strip originate from the fact that the former is developable and the latter not. Generally speaking, the ray dynamics on a curved surface is not necessarily different from that of a flat billiard, only a nonvanishing Gaussian curvature will lead to a real difference. Hence nondevelopable surfaces are of greater interest and more likely to exhibit novel effects.

\section{Calculation of geodesics} \label{sec:calcGeod}

\subsection{Local approach}
A curve $\gamma(s)$ parametrized by $s$ on a curved surface along $\vec r= \vec r\, [q_1(s),q_2(s)]$  is a geodesic if it fulfills
\begin{equation}
\label{eq:geodesic}
\frac{d^2 q^k}{ds^2} + \sum \limits_{i, j} \Gamma_{ij}^k \frac{d q^i}{ds} \frac{d q^j}{ds} = 0
\end{equation}
with $i, j, k = 1, 2$ (see Eq.~7.19b in Ref.~\cite{Nakahara}). The Christoffel symbols are given by
\begin{equation} \Gamma_{ij}^k = \frac{1}{2} \sum \limits_m g^{km} \left( \frac{\partial g_{mi}}{\partial q^j} + \frac{\partial g_{mj}}{\partial q^i} - \frac{\partial g_{ij}}{\partial q^m} \right) \,  \end{equation}
where $(g_{ij})$ is the metric tensor and $(g^{ij})$ its inverse. For the 3D M\"obius strip
\begin{equation} g_{11} = R^2 + \frac{q_2^2}{4}\, [3 + 2 \cos(q_1)] + 2 R\, q_2 \cos\left(\frac{q_1}{2}\right)  \, \end{equation}
	with the coordinates $q=(q_1,q_2)$ parametrizing the surface of a 3D M\"obius strip as defined in the main text. The other tensor elements are $g_{22} = 1$ and $g_{12} = g_{21} = 0$. 
	
\begin{figure}
\begin{center}
\includegraphics[width = 8.4 cm]{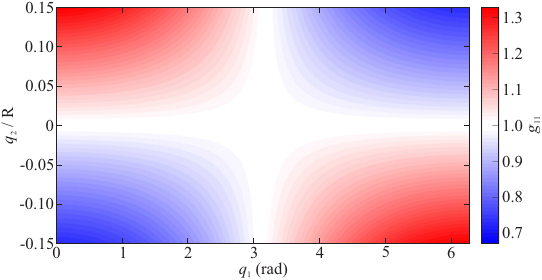}
\end{center}
\caption{\label{fig:g11}
Metric tensor element $g_{11}$ for the 3D M\"obius strip as function of the generalized coordinates $q=(q_1, q_2)$.}
\end{figure}

The element $g_{11}$ is shown as function of $(q_1, q_2)$ in Fig.~\ref{fig:g11}. Since $g_{11}$ is not constant, the 3D M\"obius strip is nondevelopable, and the Gaussian curvature of the M\"obius strip varies locally. This has important consequences for the structure of the periodic geodesics as compared to the flat M\"obius billiard as explained in Appendix~\ref{sec:PG-3Dmoebius}.	

Solving this initial conditions problem for finding the periodic geodesics evidently requires knowledge of the initial values of $q(s)$ and of $\dot{q}(s)=dq(s)/ds$. For the M\"obius strip, Eqs.~(\ref{eq:geodesic}) must be solved numerically with an algorithm such as the fourth-order Runge-Kutta method. 

Finding periodic geodesics by solving an initial conditions problem is difficult since it is \textit{a priori} unknown which initial position and velocity lead to a periodic trajectory. Hence, a large parameter space needs to be explored with high resolution. Moreover, it is difficult to judge whether a numerical solution of these equations is indeed a periodic geodesic, in view of the unavoidable numerical inaccuracies. A variational approach is simpler and more efficient. 

\subsection{Variational approach} \label{ssec:calcGeodVar}
Since a geodesic is the locally shortest path between two points --- which is essentially Fermat's principle --- we can find a geodesic segment connecting two points with coordinates $Q^{(0)}=q(0)$ and $Q^{(1)}=q(1)$ by searching the trajectory $\gamma(s)$ for which the length $L_\gamma$ given by
\begin{equation} \label{eq:Lgamma-supp}
L_{\gamma}= \int_0^1 ds \left[ \sum \limits_{i,j} g_{ij}[q(s)] \,\frac{d q_i}{ds}(s) \,\frac{d q_j}{ds}(s)  \right]^{1/2}
\end{equation}
is stationary. This formulation as a boundary condition problem no longer requires \textit{a priori} knowledge of the initial velocity $\dot{q}$. However, the two vertices $Q^{(0,1)}$ must be given beforehand and are fixed. Constructing a periodic geodesic requires calculating several such segments that are connected in a closed loop with vertices on the boundary of the M\"obius strip. Then all the vertices, including the initially fixed vertices $Q^{(0,1)}$, are varied. This will be described in more detail in the following. 

We start with the algorithm to calculate a single geodesic segment. Since it is not possible to solve analytically this infinitely-dimensional variational problem for the 3D M\"obius strip, we tackle it numerically using a discretized trajectory $\gamma = \{q(n)\}$ with $n = 1 \dots N$ where $q(1)$ corresponds to $Q^{(0)}$ and $q(N)$ to $Q^{(1)}$. The integral for $L_\gamma$ is then approximated by a sum over the $N$ discretization points using the trapezoidal rule where the derivatives of $q(s)$ in Eq.~(\ref{eq:Lgamma-supp}) are replaced by central finite differences. The trajectory $\gamma$ is stationary when all the derivatives $\partial L_\gamma / \partial q_i(n)$ vanish. Here, variation is performed perpendicular to the curve $\gamma(s)$. For this, the derivative $v_n = \partial L_\gamma / \partial q_\perp(n)$ is computed in the direction $q_\perp(n)$ which is perpendicular to the trajectory in $q$ space at discretization point $n$, implying that the two coordinates $[q_{1}(n),q_{2}(n)]$ at discretization point $n$ are not varied independently. Note that a variation tangential to the trajectory would correspond to a perturbation along the trajectory which implies a change in $s$. Therefore, as is general practice, we do not consider this variation. 

The trajectory segment is initialized with a straight line in $q$ space. The correct solution for which $\vec v$ vanishes is found using a Newton iteration, that is, $q(n) \rightarrow q(n) - d_n \, q_\perp(n)$ where $\vec{d}$ -- a $N$ component vector -- is the solution of $H \vec{d} = \vec{v}$. The matrix elements of $H$ are
\begin{equation} H_{mn} = \frac{\partial^2 L_\gamma}{\partial q_\perp(m) \, \partial q_\perp(n)} \, , \end{equation}
so $\vec{v}$ and $H$ are the first and second derivatives (gradient and Hessian), respectively, of $L_\gamma$ with respect to the $q_\perp(n)$. Since the first and last point of the segment are kept fixed, only the elements with indices in the range $2 \dots N-1$ of vectors $\vec{v}$ and $\vec{d}$ and matrix $H$ are considered in the above equations. The Newton iteration is continued till $\vec{v}$ converges to $\vec{0}$ within machine precision. 

In order to find a periodic geodesic consisting of $M$ segments, $M+1$ vertices $Q^{(m)}$ on the boundary of the M\"obius strip are chosen with values $Q_2^{(m)} = \pm W/2$ selected according to a symbolic code (see below) and the $Q_1^{(m)}$ chosen randomly in ascending order. In a first step, the $M$ individual segments are iterated as described above to find the geodesics connecting the different vertices which are kept fixed. In a second step, one step of the Newton iteration for all discretization points including all of the vertices $Q^{(m)}$ is calculated, where the vertices are only allowed to vary in the $q_1$ direction to keep them on the boundary of the M\"obius strip. Steps one and two are repeated till the trajectory becomes stationary within machine precision. To ensure periodicity, $Q^{(0)}$ and $Q^{(M)}$ are initialized such that they represent the same point in configuration space, that is, $Q_1^{(M)} = Q_1^{(0)} + 2 \pi n_l$ and $Q_2^{(M)} = (-1)^{n_l} Q_2^{(0)}$ where $n_l$ is the number of revolutions around the M\"obius strip, and the same variation is applied to both of them. 

A large number of random initial conditions for the vertices $Q^{(m)}$ was tested for each considered symbolic code since the algorithm only converges to a physical trajectory if the initial conditions are sufficiently close to a correct solution. A vanishing gradient $\vec{v}$ proves that the calculated trajectories are numerically exact geodesics. Typically, $10,000$ discretization points per revolution were used in the calculations, and it was checked that increasing the number of discretization points did not significantly change the length $L_\gamma$ or the shape of the trajectory. 

Furthermore, it was verified that the law of elastic reflection is respected at all vertices, i.e., that the ingoing angle is equal to the outgoing angle. It should be noted that the law of elastic reflection is not \textit{a priori} assumed in any way during the numerical calculations of the periodic geodesics but results when a correct solution is found. In addition, several periodic geodesics were verified independently by solving the equations (\ref{eq:geodesic}) with the correct initial conditions.

\section{Periodic geodesics of the 3D M\"obius strip} \label{sec:PG-3Dmoebius}

\begin{figure*}
\begin{center}
\includegraphics[width = 15 cm]{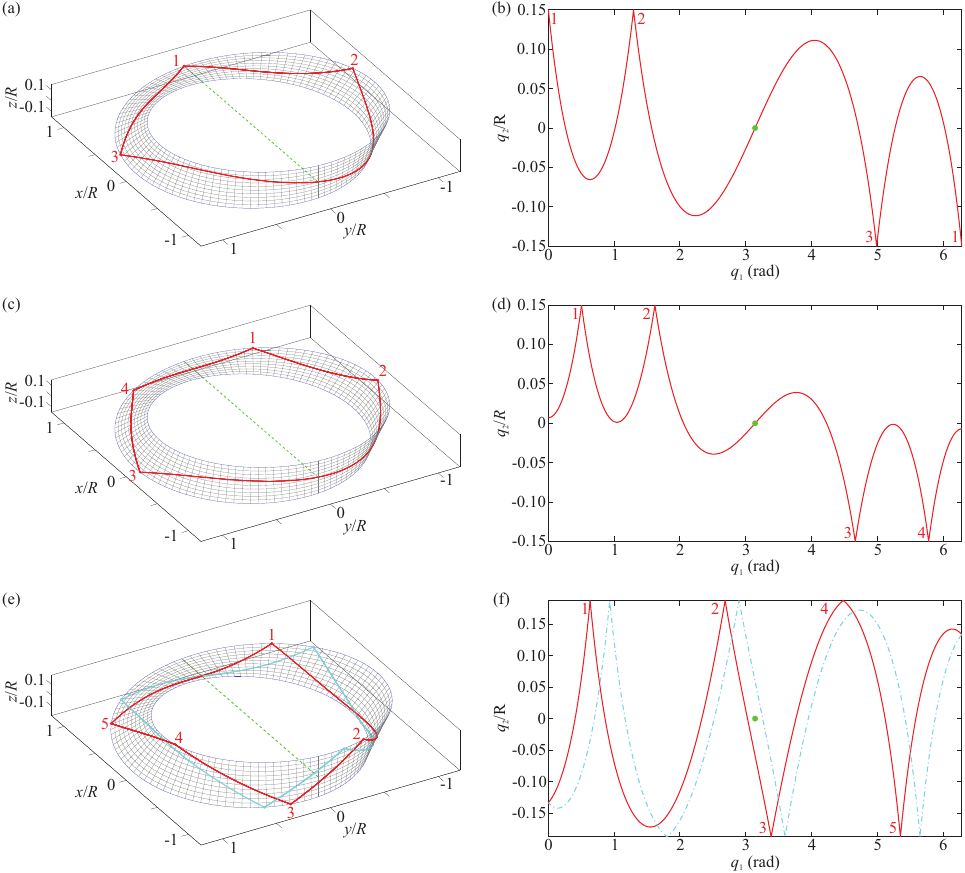}
\end{center}
\caption{\label{fig:PerGeoSM}
Further examples of periodic geodesics (red solid lines) in configuration (a, c, e) and $q$ space (b, d, f). The green dots and the dashed green line indicates the symmetry axis. (a, b) Periodic geodesic $(3, 1, sos)$ with length $L = 6.53 R$ on a M\"obius strip with aspect ratio $W/R = 0.3$. (c, d) Periodic geodesic $(4, 1, sos^2)$, called 4a in the following, with length $L = 6.66 R$ on a M\"obius strip with $W/R = 0.3$. (e, f) Periodic geodesic $(5, 1, so^3s)$ with length $L = 6.67 R$ on a M\"obius strip with $W/R = 0.375$. Its symmetry partner obtained by rotation around the $x$-axis (green dashed line) is indicated as cyan dash-dotted line.}
\end{figure*}

In this Appendix, additional examples of periodic geodesics of the 3D M\"obius strip are shown and their structure is compared to the periodic orbits in the flat M\"obius billiard and conventional ring cavities. 

Figure~\ref{fig:PerGeoSM} shows three typical examples. Some periodic geodesics like those in Figs.~\ref{fig:PerGeoSM}(a-d) are symmetric with respect to a rotation of $180^\circ$ around the $x$-axis, which is the symmetry axis of the M\"obius strip and indicated as green dashed line in the configuration space plots. These trajectories all pass through the point $q = (\pi, 0)$ indicated as green dot in the $q$ space plots. Other orbits such as the one shown in Figs.~\ref{fig:PerGeoSM}(e, f) are not, and they transform to a distinct periodic geodesic under the symmetry rotation by 180$^{\circ}$. 

The periodic geodesics are labeled by indices $(n_t, n_l)$ and a symbolic code, where $n_t$ is the number of vertices at the boundary and $n_l$ is the number of revolutions around the M\"obius strip. The symbolic code consists of $n_t$ letters that connect vertices with a different or same sign of $q_2$, i.e., reflection points on Opposite boundaries or the Same boundary, though the notion of ``same'' and ``opposite'' side only makes sense from a local perspective given that the M\"obius strip actually has only one boundary\footnote{The labels are not unambiguous since several distinct periodic geodesics with the same indices and code can exist.}. The $(11, 1, o^{11})$ orbit in Figs.~2(c, e) of the main text, for example, has a zig-zag structure where all segments connect opposite sides, whereas the $(5, 1, s^2os^2)$ orbit in Figs.~2(d, f) of the main text starts with two segments connecting the same side, followed by a segment connecting to the opposite side after half a round trip followed by two segments connecting the same side again. 

The existence of specific periodic geodesics depends on the aspect ratio $W/R$. For example, the $(5, 1, so^3s)$ geodesic in Figs.~\ref{fig:PerGeoSM}(e, f) does not exist for smaller aspect ratios such as $W/R = 0.3$. As the aspect ratio is decreased and the strip becomes narrower, the segment connecting vertices $1$ and $2$ eventually touches the lower boundary $q_2 = -W/2$ and the geodesic hence ceases to exist. In contrast, the periodic orbits of the flat M\"obius billiard exist independently of its aspect ratio. 

The three periodic geodesics in Fig.~\ref{fig:PerGeoSM} all have structures that cannot be found for periodic orbits of the flat M\"obius billiard. First, all of them feature segments that connect to the same boundary (code $s$), however such segments are impossible in the flat M\"obius billiard. Furthermore, some periodic geodesics such as the $(4, 1, sos^2)$ in Figs.~\ref{fig:PerGeoSM}(c, d) have $n_t + n_l$ odd, whereas it must always be even for the flat M\"obius. First, these differences can be directly attributed to the fact that the 3D M\"obius strip is nondevelopable and thus has a nonvanishing Gaussian curvature, implying that some parts of the boundary are convex and thus permit $s$ trajectory segments, whereas the the straight boundary of the flat M\"obius billiard does not allow them. Second, $n_t + n_l$ must be even in the flat M\"obius since $\dot{y}$ can only change its sign at a reflection but stays constant in between, whereas $\dot{q}_2$ changes continuously during the propagation on the curved surface of the 3D M\"obius strip in addition to the sign changes at the vertices. 

Another difference is that the POs of the flat M\"obius billiard are all part of a continuous family whereas the periodic geodesics of the 3D M\"obius strip are all isolated, with one exception given below. The existence of families of POs is due to the continuous translational symmetry in $x$ direction of the flat M\"obius billiard. While the 3D M\"obius strip has a similar periodic boundary condition as the flat M\"obius billiard, the metric tensor $g_{ij}$ does not exhibit a continuous translational symmetry, see Fig.~\ref{fig:g11}. The periodic geodesics are isolated due to this reduction of symmetry. 

\begin{figure}
\begin{center}
\includegraphics[width = 8.4 cm]{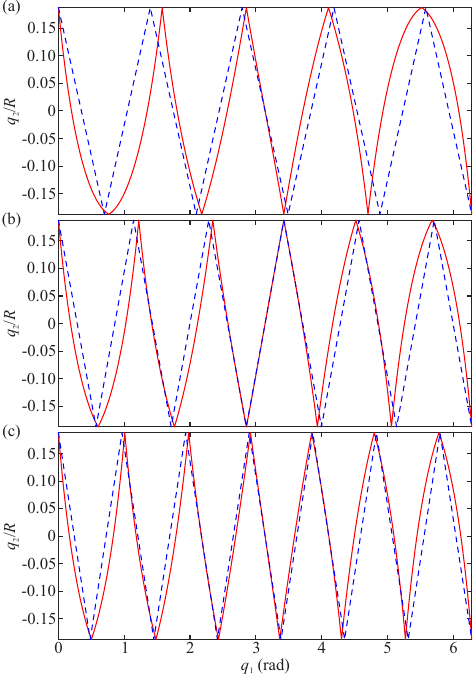}
\end{center}
\caption{\label{fig:DevPOs}
Periodic geodesics of $(n_t, 1, o^{n_t})$ type (red solid lines) on a 3D M\"obius strip with apsect ratio $W/R = 0.375$. They resemble the $(n_t, 1)$ periodic orbits of the flat M\"obius billiard which are indicated as blue dashed lines. (a) Geodesic $(9, 1, o^9)$ with length $L = 7.03 R$. The length of the $(9, 1)$ PO is $7.13 R$. (b) Geodesic $(11, 1, o^{11})$ with length $L = 7.44 R$. The length of the $(11, 1)$ PO is $7.52 R$. (c) Geodesic $(13, 1, o^{13})$ with length $L = 7.89 R$. The length of the $(13, 1)$ PO is $7.95 R$.}
\end{figure}

There is one series of periodic geodesics, the $(n_t, 1, o^{n_t})$ orbits with $n_t$ odd, that resembles the $(n_t, 1)$ periodic orbits of the flat M\"obius billiard. Three examples of such periodic geodesics are shown as red lines in Fig.~\ref{fig:DevPOs}. The analog POs of the flat M\"obius billiard are indicated as blue dashed lines, where the coordinate $x$ ($y$) of the flat M\"obius billiard is identified with $q_1$ ($q_2$). As the number of reflections $n_t$ increases, the segments of the periodic geodesics become less and less curved in $q$ space and increasingly resemble the straight segments of the POs of the flat M\"obius billiard, where the segments in the region around $q_1 = 0$, $2 \pi$ exhibit generally more curvature than those in the region around $q_1 = \pi$. Also the lengths of the periodic geodesics become closer to the lengths of the corresponding POs in the flat M\"obius. 

While the $(n_t, 1, o^{n_t})$ periodic geodesics slowly converge towards the corresponding POs of the flat M\"obius billiard for increasing $n_t$, some differences remain. First, these periodic geodesics only exist for $n_t \geq 9$ for the aspect ratio $W/R = 0.375$ considered here ($n_t \geq 11$ for $W/R = 0.3$) because the concave parts of the boundary do not allow some of the longer segments needed for smaller $n_t$. Second, the periodic geodesics remain isolated since the 3D M\"obius strip lacks the translational symmetry of its flat counterpart. 

These two effects can be understood by considering the metric tensor. For the flat M\"obius billiard, $g_{ij} = \delta_{ij}$, and hence all geodesics are straight lines. For the 3D M\"obius strip, $g_{22} = 1$ as well, but $g_{11}$ deviates from unity (see Fig.~\ref{fig:g11}) implying that the geodesics are curved in $q$ space. However, the more the velocity vector $\dot{q}$ of a trajectory becomes parallel to the $q_2$ direction and thus $\dot{q}_1$ decreases, which is necessarily the case when $n_t$ increases, the less important the contribution of $g_{11}$ is, which results in less curvature. Therefore, the segments of the periodic geodesics resemble more and more straight lines as $n_t$ increases. Moreover, the deviation of $g_{11}$ from $1$ is smallest in the region around $q_1 = \pi$, and hence the trajectory segments are less curved there. 

A special case are the $(2, 0, o^2)$ geodesics that bounce back and forth perpendicularly between the sides of the M\"obius strip, in analogy to the bouncing-ball orbits of the flat M\"obius billiard shown in Fig.~\ref{fig:POplaneMoebius}(f). They are the only periodic geodesics that form a continuous family, that is, any trajectory with $q_1 = \mathrm{const.}$ belongs to this family. Since $\dot{q}_1 = 0$, the spatially varying tensor element $g_{11}$ has no effects for these geodesics, whereas $g_{22} = 1$ is homogeneous. Thus, these periodic geodesics have the same properties as their counterparts in the flat M\"obius billiard. 

While the flat M\"obius billiard features POs without reflections at the boundaries, i.e., the $(0,n_l)$ orbits in Fig.~\ref{fig:POplaneMoebius}(b), such periodic geodesics were not found for the 3D M\"obius strip. In particular, trajectories with $q_2 = \mathrm{const.}$ such as the one shown in Fig.~\ref{fig:geodesique}(a) of the main text are not geodesics since $g_{11} \neq 1$, and hence $\dot{q}_2\neq 0$ for almost all points of a geodesic. Therefore, at least one reflection at the boundary is needed to compensate the change of sign of $\dot{q}_2$ after one round trip due to the periodic boundary conditions with inversion of $q_2$. 

It is also interesting to compare the M\"obius strip with conventional whispering gallery resonators exhibiting azimuthal rotational symmetry such as rings, spheres or toroids. Such resonators are of interest since they feature modes with ultra-high quality ($Q$) factors due to the existence of whispering gallery trajectories that propagate along the outer resonator wall with a constant angle of incidence above the critical angle for total internal reflection. High-$Q$ modes exist even for small deformations of the rotational symmetry that break angular momentum conservation. 

Evidently, the 3D M\"obius strip does not have an azimuthal symmetry that would preserve the angular momentum. More importantly, whispering gallery type trajectories, which would have a symbolic code $s^{n_t}$, cannot exist since parts of the boundary are concave and thus locally prohibit $s$-type segments. Moreover, the boundary of the M\"obius strip is not a geodesic itself. Hence, all periodic geodesics must feature at least one $o$-segment that crosses the strip. In many cases an $o$-segment passing through or near the symmetry point $q = (\pi, 0)$ is found. As a consequence, the 3D M\"obius strip cannot support whispering gallery modes, and the intensity distributions of its resonant modes typically show the same crossing behavior [see Fig.~\ref{fig:num-resonance}(d) in the main text and Figs.~\ref{fig:mode3}(d) and \ref{fig:mode8}(d) below]. Nonetheless, resonances with high quality factors exist since many periodic geodesics such as the ones shown above are confined by total internal reflection at all reflections.

\section{3D FDTD simulations} \label{sec:3dFDTD}

In order to better understand the relation between the geodesics and the resonant modes of the M\"obius strip, the spectrum as well as the field distributions of several modes are calculated for a M\"obius strip with $R = 10~\mu$m, $W = 3~\mu$m, $h = 150$~nm, and $n = 1.515$. Two examples of mode patterns are presented here in addition to the one shown in Fig.~\ref{fig:num-resonance} of the main text. The calculation of the length spectrum is explained as well.

\subsection{Spectrum of the M\"obius strip}

\begin{figure}[bt]
\begin{centering}
\includegraphics[width=8cm]{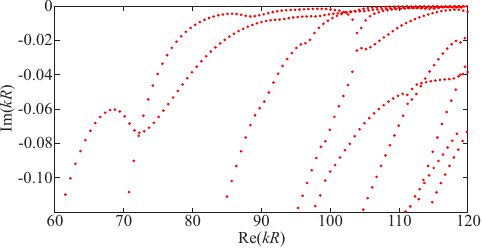}
\par\end{centering}
\caption{\label{fig:SpectrumFDTD}
Real versus imaginary parts of the eigenwavenumbers (i.e, spectrum) calculated with 3D FDTD simulations for a M\"obius strip with $R = 10~\mu$m, $W = 3~\mu$m, $h = 150$~nm, and $n = 1.515$. Only the modes symmetric with respect to the $x$-axis are shown.}
\end{figure}

To obtain the spectrum of a M\"obius strip, a 3D finite difference time domain (FDTD) simulation is performed with a temporally short (spectrally broad) excitation. The time evolution of the field at one point in the M\"obius strip is recorded. To get an accurate estimate of the frequencies and lifetimes from the recorded time data, a spectral estimation technique known as Prony's method is used \cite{Marple}.
The calculated spectrum in Fig.~\ref{fig:SpectrumFDTD} features several series of roughly equidistant resonant modes. 

Since the M\"obius strip exhibits a two-fold rotation symmetry with respect to the $x$-axis ($C_2$ symmetry), its modes are either symmetric or antisymmetric with respect to this rotation. The spectrum with symmetric modes in Fig.~\ref{fig:SpectrumFDTD} is calculated using an excitation dipole in the middle of the flat part of the strip at $(R, 0, 0)$ with electric field parallel to the $x$-axis. Antisymmetric modes (not shown) are calculated as well by displacing the excitation dipole away from the symmetry axis. The symmetric and antisymmetric modes are almost degenerate, with a relative deviation of typically $2\cdot10^{-2}$ in $\mathrm{Im}(kR)$ and an even smaller relative deviation of about $2\cdot10^{-7}$ in $\mathrm{Re}(kR)$. It should be noted that a $C_2$ symmetry does not imply degeneracies. 

In addition we performed simulations with posts which were positioned as in the experiments. The resulting series of modes barely differed from those obtained with no posts. However, the losses slightly increased because a small fraction of modes propagating within the M\"obius strip was scattered out by the posts. In summary, the posts do not significantly impact the structure of the spectrum and hence our conclusions about the relation to periodic geodesics remain valid.

\subsection{Mode patterns and geodesics} \label{ssec:modes}

\begin{figure}[bt]
\begin{centering}
\includegraphics[width=8cm]{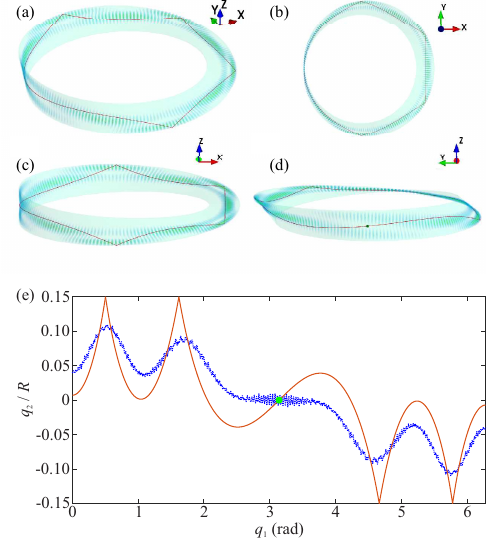}
\par\end{centering}
\caption{\label{fig:mode3}
(a)-(d) Different views of the same mode with $\mathrm{Re}(kR) = 77.843$ and $\mathrm{Im}(kR) = -0.0494$ ($Q = 788$). $|\vec{E}|^2$ is shown in false colors. A geodesic of type $(4, 1, sos^2)$ with length $6.656 R$ is superimposed as red lines. The green dot in panel (d) indicates the intersection of the $x$-axis with the vertical section of the M\"obius strip. (e) Representation of the geodesic (red solid line) as function of $q_{1, 2}$. The blue dotted line is the mean position of the mode calculated from Eq.~(\ref{eq:mean-position}) in the main text. The green dot indicates the symmetry axis.}
\end{figure}

\begin{figure}[bt]
\begin{centering}
\includegraphics[width=8cm]{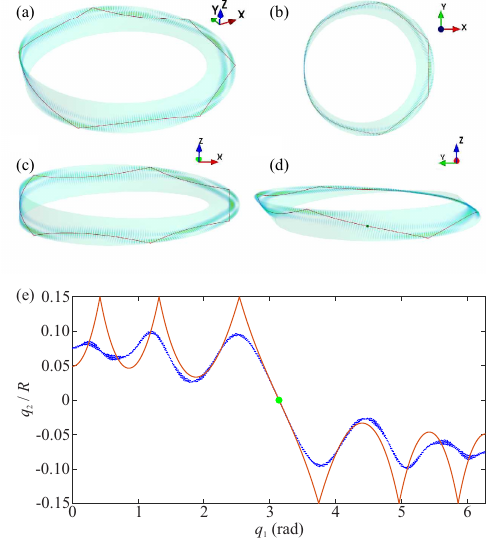}
\par\end{centering}
\caption{\label{fig:mode8}
(a)-(d) Different views of the same mode with $\mathrm{Re}(kR) = 109.260$ and $\mathrm{Im}(kR) = -0.00021$ ($Q = 265,370$). $|\vec{E}|^2$ is shown in false colors. A geodesic of type $(6, 1, s^2os^3)$ with length $6.779 R$ is superimposed as red lines. The green dot in panel (d) indicates the intersection of the $x$-axis with the vertical section of the M\"obius strip. (e) Representation of the geodesic (red solid line) as function of $q_{1, 2}$. The blue dotted line is the mean position of the mode calculated from Eq.~(\ref{eq:mean-position}) in the main text. The green dot indicates the symmetry axis.}
\end{figure}

To calculate the field distribution of a particular resonance, a narrow band excitation at the targeted frequency is used so that only the targeted mode is excited in the FDTD simulation. Two examples in addition to Fig.~\ref{fig:num-resonance} in the main text are shown in Figs.~\ref{fig:mode3} and \ref{fig:mode8}. In contrast to the mode shown in Fig.~\ref{fig:num-resonance}, which exhibits only a single transverse excitation, the modes in Figs.~\ref{fig:mode3} and \ref{fig:mode8} have a second-order transverse excitation. The three modes belong to different branches of the spectrum, respectively. 

The mode in Fig.~\ref{fig:mode3} shows good agreement with a periodic geodesic of type $(4, 1, sos^2)$, geodesic 4a shown in Fig.~\ref{fig:PerGeoSM}(c, d). The agreement of the mode in Fig.~\ref{fig:mode8} with geodesic $(6, 1, s^2os^3)$ is excellent. These examples confirm that some modes of the M\"obius strip are closely related to periodic geodesics of various types, though not all modes show a clear connection to a single periodic geodesic. It furthermore appears that better agreement is found for modes with higher excitation frequency $\mathrm{Re}(kR)$, that is, in the semiclassical limit. 

\subsection{Length spectrum} \label{ssec:Lspect}
Figure~\ref{fig:num-spectre} shows the Fourier transform of the numerically calculated spectrum (cf.\ Fig.~\ref{fig:SpectrumFDTD}). It is referred to as \emph{length spectrum} since it peaks at the lengths of the underlying periodic orbits or geodesics. Since the effective refractive index of the simulated M\"obius strip varies significantly as function of frequency between $n_\mathrm{eff} \simeq 1.12$ and $1.27$ in the range of $\mathrm{Re}(kR) = 60$--$120$, the Fourier transform is slightly modified in order to account for this dispersion. Following Ref.~\cite{bittner2}, the Fourier transform is calculated as
\begin{equation} \hat{\rho}(\ell) = \sum \limits_j \exp\{ -i k_j  \ell n_\mathrm{eff}[\mathrm{Re}(k_j)] \} \end{equation}
where the $k_j$ are the eigenwavenumbers of the resonances and the refractive index in the exponential was replaced by the effective refractive index at the respective resonance frequency. A total of $628$ symmetric modes in the range of $60 \leq \mathrm{Re}(kR) \leq 120$ and with $\mathrm{Im}(kR) \geq -0.60$ were used in the calculation. Figure~\ref{fig:num-spectre} shows $|\hat{\rho}|$ as function of the geometric length $\ell$.

\section{Effective refractive index}

The effective index approximation assumes that the 3D (scalar) Helmholtz equation
\begin{equation}\label{eq:helmholtz-scal}
(\Delta + n^2k^2)\,\psi=0
\end{equation}
can be approximated by the 2D Helmholtz equation
\begin{equation}
\Delta_s\psi_s+n_\mathrm{eff}^2k^2\psi_s=0
\end{equation}
where $\Delta_s$ is the Laplace operator on the curved surface and $\psi_s(q_1,q_2)$ the wave function on the surface. Then the effective index $n_\mathrm{eff}$ includes the influence of the finite thickness of the resonator in the $q_3$ direction perpendicular to the surface. %

This Appendix deals with two issues. In Appendix~\ref{ssec:derivation}, we derive the effective index for a curved layer and show that, for the range of parameters considered in the experiments, there is no difference with a flat layer. In Appendix~\ref{ssec:exp}, we investigate the experimental value of the group refractive index $n_g$ and its uncertainty, which yields the value $n_g=1.58\,\pm\,0.05$.

\subsection{Derivation} \label{ssec:derivation}

The effective index approximation is usually computed for flat layers. However, in the ($q_1$,\,$q_3$) plane, a section of the M\"obius strip locally looks like a ring, with a local radius of curvature varying from $R$ to $\infty$ (i.e., a flat layer). Following Ref. \cite{jap-indice-courbe}, we extend the derivation of the effective index approximation to a curved layer, which is assumed to be a segment of a ring of inner radius $a$ and outer radius $b$. In correspondence with experiments, we use $b=50~\mu$m, $b-a = 1~\mu$m, and a bulk refractive index $n=1.515$ for the photoresist \cite{indice}. 

First the eigenfunctions $\phi$ of the ring are calculated based on the ansatz
\begin{equation} \begin{array}{ccl} r < a & : & \phi(r,\theta)=A J_m(k r) e^{i m \theta} \\ \\
a < r < b & : & \phi(r,\theta)=J_m(n k r) e^{i m \theta} + B Y_m(n k r) e^{i m \theta} \\ \\
b < r & : & \phi(r,\theta)=C H_m^{(1)}(k r) e^{i m \theta}
\end{array} \end{equation}
where $J_m$ and $Y_m$ are the Bessel functions of the first and second kind, respectively, $H_m^{(1)}$ is the Hankel function of the first kind, and their order $m$ is an integer. $A$, $B$, and $C$ are constants, and $r$, $\theta$ are the standard circular coordinates. We consider the polarization of the electric field to be parallel to the layer. Applying the continuity conditions at the boundaries $r=a$ and $r=b$, we get
\begin{eqnarray}
\frac{\frac{J'_m(ka)}{J_m(ka)}\,J_m(nka)-nJ'_m(nka)}
{nY'_m(nka)-Y_m(nka)\,\frac{J'_m(ka)}{J_m(ka)}}
=\nonumber \\
\frac{\frac{H'_m(kb)}{H_m(kb)}\,J_m(nkb)-nJ'_m(nkb)}
{nY'_m(nkb)-Y_m(nkb)\,\frac{H'_m(kb)}{H_m(kb)}}
\label{eq:k-anneau} \end{eqnarray}
where the prime denotes the derivative. For each value of $m$, there are several eigenwavenumbers $k$ which correspond to different excitations in the direction $q_3$. The effective index is defined via
\begin{equation}
m = n_\mathrm{eff}kr_0
\end{equation}
where $r_0$ is the radial center of mass of the wave function in the ring. In the frequency range corresponding to the experiments, the wave functions are well centered in the ring, so we assume $r_0$ in the middle, $r_0=(a+b)/2$. We calculate the $k$ eigenwavenumbers for various $m$ values based on Eq.~(\ref{eq:k-anneau}) and then plot
\begin{equation}
n_\mathrm{eff}=\frac{m}{k\,r_0} \, .
\end{equation}

\begin{figure}[tb]
\begin{center}
\includegraphics[width = 1\linewidth]{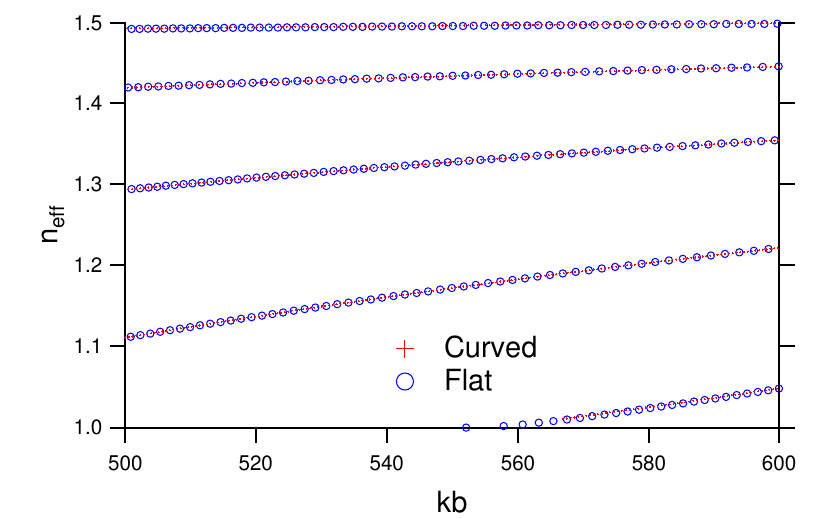}
\end{center}
\caption{Effective refractive index for a flat layer (Eq.~(4) in Ref.~\cite{PRE-trace1}) and a curved layer for $n=1.515$, an outer radius $b=50~\mu$m, and an inner radius $a=49~\mu$m in the experimental wavelength range $\lambda=2\pi/k=560$~nm to $\lambda=600$~nm. The electric field is assumed parallel to the layer.}
\label{fig:indice-dispersion}
\end{figure}

Figure~\ref{fig:indice-dispersion} shows the comparison of the effective index for a flat and a curved layer in the range of $k$ corresponding to the experiments. The effective indices for flat and curved layers agree excellently for the parameter range considered here, that is, the curvature causes a negligible change of $n_\mathrm{eff}$ compared to the flat case even for the smallest radius of curvature, $50~\mu$m, found in the M\"obius strip. Consequently we can assume that the effective index is constant throughout the M\"obius strip (i.e., independent of the local curvature) and equal to that of a flat layer with the same thickness.

\subsection{Experiments} \label{ssec:exp}

As discussed in the main text, the Fourier transform of the laser spectrum exhibits peaks at the optical length(s) of the underlying periodic orbit(s). The geometrical length of these trajectories is obtained by dividing by the group refractive index (see Ref.~\cite{APL-FP3D}). It is the sum of the effective index and contributions from the material dispersion of the bulk index $n$ and modal dispersion due to $n_{\mathrm{eff}}$ approximation,
\begin{equation}
\label{eq:indice-groupe}
n_g=n_\mathrm{eff}+\left.k\frac{\partial n}{\partial k}\right|_\mathrm{mat}+\left.k\frac{\partial n}{\partial k}\right|_\mathrm{eff} \, .
\end{equation}
The bulk refractive index is n=1.515 $\pm$ 0.005 \cite{indice}. Its dispersion is very small, thus we assume that $n$ is constant when calculating $n_\mathrm{eff}$. However, $k$ is large, hence the term $\left.k\frac{\partial n}{\partial k}\right|_\mathrm{mat}$ cannot be neglected. Following Ref.~\cite{APL-FP3D}, it is inferred from the laser spectra of cuboids since their lasing modes are located on a well-known periodic orbit in the bulk of the cavity. From these data we estimate the material dispersion as $\left.k\frac{\partial n}{\partial k}\right|_\mathrm{mat}=$ 0.05 $\pm$ 0.04. 

The effective index and its dispersion are calculated as discussed in the previous section. In the frequency range of interest and for a $1~\mu$m thick strip, there are four to five branches (excitations in the $q_3$ direction) for each polarization with significantly different indices and dispersion values. Since the spectrum in Fig.~\ref{fig:manips-spectre}(a) exhibits a single series of resonances, we consider only the branch with the highest effective index which has the best confinement factor and thus the highest laser gain. The corresponding effective index is $n_\mathrm{eff}=1.49$ and its dispersion equals $\left. k\frac{\partial n}{\partial k}\right|_\mathrm{eff}=0.037\,\pm\,0.002$. Finally, adding the three terms in Eq.~(\ref{eq:indice-groupe}) yields a group index of $n_g=1.58\,\pm\,0.05$.

\section{Additional experiments} \label{sec:addExp}

\subsection{Photographs}

\begin{figure}[bt]
\begin{center}
\includegraphics[width = 0.8\linewidth]{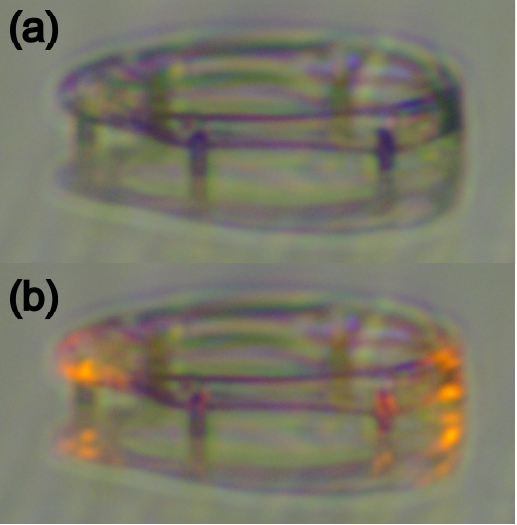}
\end{center}
\caption{Photographs of a M\"obius strip illuminated with low intensity white light (a) without pump laser and (b) with pump. We can see the cavity and its reflection on the glass substrate.  The resist IP-G780 is transparent.
The green pump light is removed by a notch filter, while the laser emission is yellow. }
\label{fig:photo-pompage}
\end{figure}

Figure~\ref{fig:photo-pompage} shows a lasing M\"obius strip cavity registered with a Ueye camera equipped by a zoom lens (by Navitar). The horizontal section of the M\"obius surface is on the left and the vertical section on the right. Since the photoresist is transparent, only the edges of the M\"obius strip as well as the pylons are seen as dark lines in the photographs.

\subsection{Influence of the strip thickness} \label{ssec:thickness}

\begin{figure}[tb]
\begin{center}
\includegraphics[width = 1\linewidth]{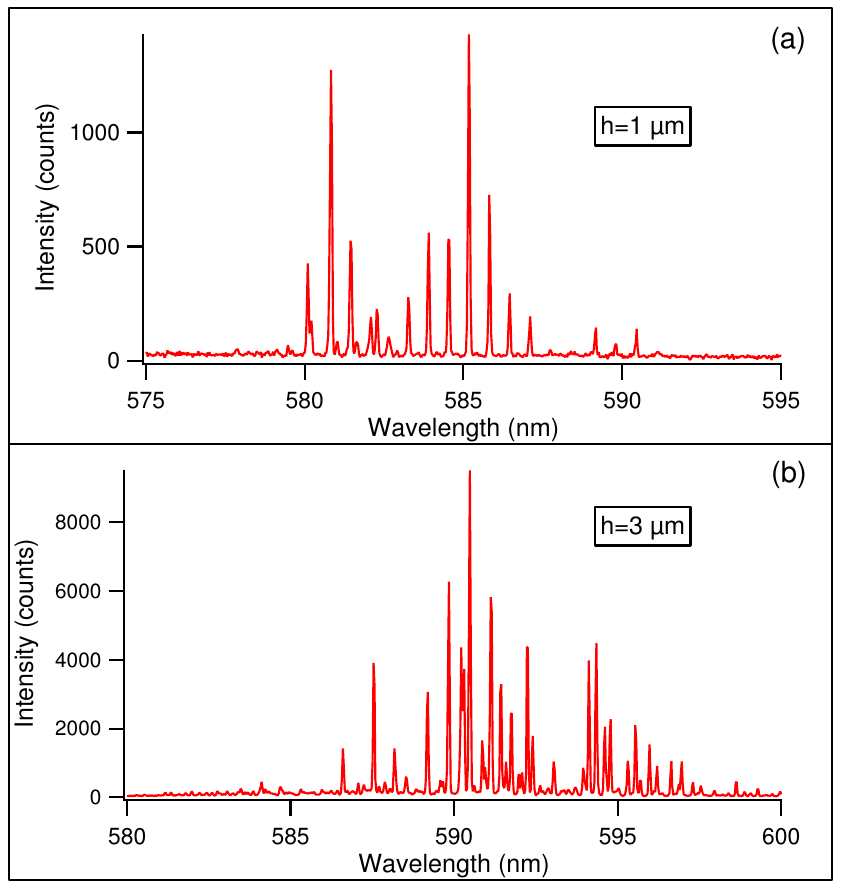}
\end{center}
\caption{Experimental spectra of M\"obius strip microlasers for $R=50~\mu$m and $W = 15~\mu$m with (a) $h=1~\mu$m, and (b) $h=3~\mu$m. All experimental parameters are identical for (a) and (b), in particular the pump intensity, which is four times the laser threshold of the microcavity in (a).}
\label{fig:epaisseur}
\end{figure}

Experiments were performed with M\"obius strip microlasers of thickness $h=1$, $2$, $3$, and $5~\mu$m, where $R=50~\mu$m and $W=15~\mu$m. For $h=1~\mu$m, the spectrum features a single series of resonances, independent of the pump intensity as shown in Fig.~\ref{fig:epaisseur}(a). For $h=3~\mu$m the spectrum exhibits a single series just above threshold, but becomes more complex as the pump intensity increases [see Fig.~\ref{fig:epaisseur}(b)]. However, its Fourier transform is peaked at the same optical length independently of the pump intensity. For $h=5~\mu$m thickness, in contrast, the spectrum and its Fourier transform are more complex, which means that the analysis using the effective index approximation is probably not valid. In Fig.~\ref{fig:manips} of the main text, only cavities with $h=1~\mu$m are considered to ensure that the effective index approximation remains valid.

\end{document}